\documentclass[journal]{IEEEtran}

\ifCLASSINFOpdf
\else
   \usepackage[dvips]{graphicx}
\fi
\usepackage{url}
\usepackage{cite}
\usepackage{color}
\usepackage{multirow}
\usepackage{booktabs}
\usepackage{subcaption}
\usepackage{amsmath}
\usepackage{caption}
\usepackage{wrapfig}
\usepackage{subcaption}
\usepackage[table]{xcolor}
\usepackage{subcaption}
\usepackage[export]{adjustbox}
\usepackage{xcolor}
\usepackage[colorlinks = true,
            linkcolor = blue,
            urlcolor  = magenta,
            citecolor = blue,
            anchorcolor = blue]{hyperref}
\providecommand{\gz}[1]{\textcolor{black}{{#1}}}
\providecommand{\zd}[1]{\textcolor{black}{{#1}}}
\usepackage{graphicx}

\newcommand{\fref}[1]{Fig.~\ref{#1}}

\begin{document}

\title{MusicHiFi: Fast High-Fidelity Stereo Vocoding}
\author{Ge Zhu$^{1, 2*}$, Juan-Pablo Caceres$^{2}$, Zhiyao Duan$^{1}$, and Nicholas J. Bryan$^{2}$ \\
\vspace{2mm}
        $^{1}$ University of Rochester \;\;\; $^{2}$ Adobe Research 
\thanks{This work is done during an internship at Adobe Research. 
Correspondence to: Ge Zhu (gzhu6@ur.rochester.edu) and Nicholas J. Bryan (njb@ieee.org).}}

\maketitle

\begin{abstract}
Diffusion-based audio and music generation models commonly perform generation by constructing an image representation of audio (e.g., a mel-spectrogram) and then convert it to waveform using a phase reconstruction model or vocoder. Typical vocoders, however, produce monophonic audio at lower resolutions (e.g., 16-24 kHz), which limits their usefulness. We propose MusicHiFi --- an efficient high-fidelity stereophonic vocoder. Our method employs a cascade of three generative adversarial networks (GANs) that convert low-resolution mel-spectrograms to audio, upsamples to high-resolution audio via bandwidth extension, and upmixes to stereophonic audio. Compared to past work, we propose 1) a unified GAN-based generator and discriminator architecture and training procedure for each stage of our cascade, 2) a new fast, near downsampling-compatible bandwidth extension module, and 3) a new fast downmix-compatible mono-to-stereo upmixer that ensures the preservation of monophonic content in the output. We evaluate our approach using objective and subjective listening tests and find our approach yields comparable or better audio quality, better spatialization control, and significantly faster inference speed compared to past work. Sound examples are at \url{https://MusicHiFi.github.io/web/}.
\end{abstract}


\IEEEpeerreviewmaketitle

\section{Introduction}
Recent generation methods offer an exciting new way to create media content. Diffusion models~\cite{ho2020denoising}, in particular, have shown great promise for fast, high-quality image generation~\cite{ramesh2022hierarchical, rombach2022high} and are rapidly being adopted for audio and music generation~\cite{Forsgren_Martiros_2022, liu2023audioldm, huang2023noise2music,hawthorne2022multi,chen2023musicldm, wu2023music, wang2024audit, novack2024ditto}. When used for audio~\cite{hawthorne2022multi, huang2023noise2music,liu2023audioldm,chen2023musicldm}, diffusion models are commonly used to generate an image representation of audio (e.g. mel-spectrogram) and then a phase reconstruction model or vocoder is used to convert to \zd{waveform}. This two-stage cascaded approach has been shown beneficial~\cite{huang2023noise2music}, but typically generates low-resolution audio (e.g., mono 16-24 kHz) via a diffusion~\cite{kong2020diffwave} or generative adversarial network (GAN) vocoder~\cite{kumar2019melgan, kong2020hifi, lee2023bigvgan}.

To improve diffusion-based generation quality for audio and music, a high-fidelity vocoder is essential. 
Current methods to do so largely come in two forms: bandwidth extension (expansion) and mono-to-stereo upmixing. 
Bandwidth extension (BWE) increases the frequency range or bandwidth of an audio signal and can be achieved using a source-filter model~\cite{iser2008bandwidth}, time-domain~\cite{su2021bandwidth,han2022nu,zhang2021wsrglow}, or spectral-domain~\cite{eskimez2019speech,kumar2020nu,hu2020phase,mandel2023aero,liu2023audiosr} neural network. Mono-to-stereo (M2S) upmixing converts a single-channel audio signal into spatialized (left and right) channels. 
M2S can be achieved via decorrelation~\cite{jot2003spatial} or a recent parametric stereo method~\cite{serra2023mono} that describes audio as a single-channel plus parameters that \zd{approximately} characterizes the stereo image~\cite{schuijers2004low}.

\begin{figure}[t!]
\centering
\begin{subfigure}[b]{0.45\textwidth}
\centering
\includegraphics[width=0.82\linewidth]{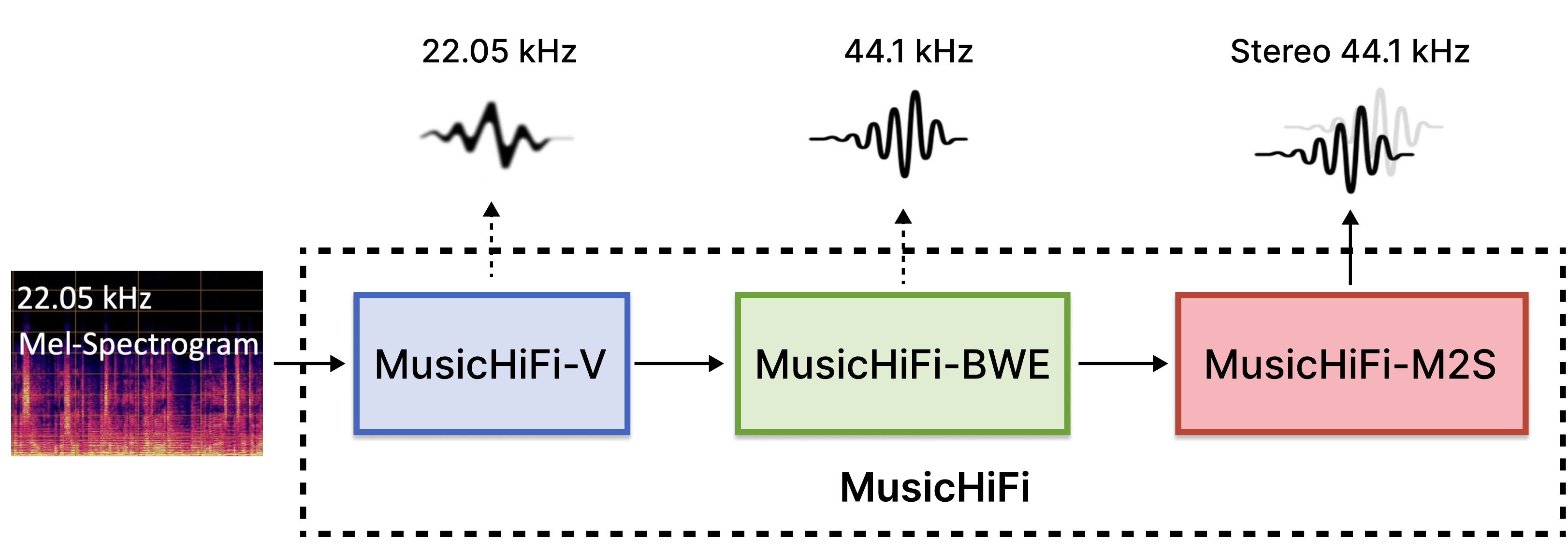}
   \caption{}
\label{subfig:diag1} 
\end{subfigure}
\begin{subfigure}[b]{0.5\textwidth}
\centering
\includegraphics[width=0.95\linewidth]{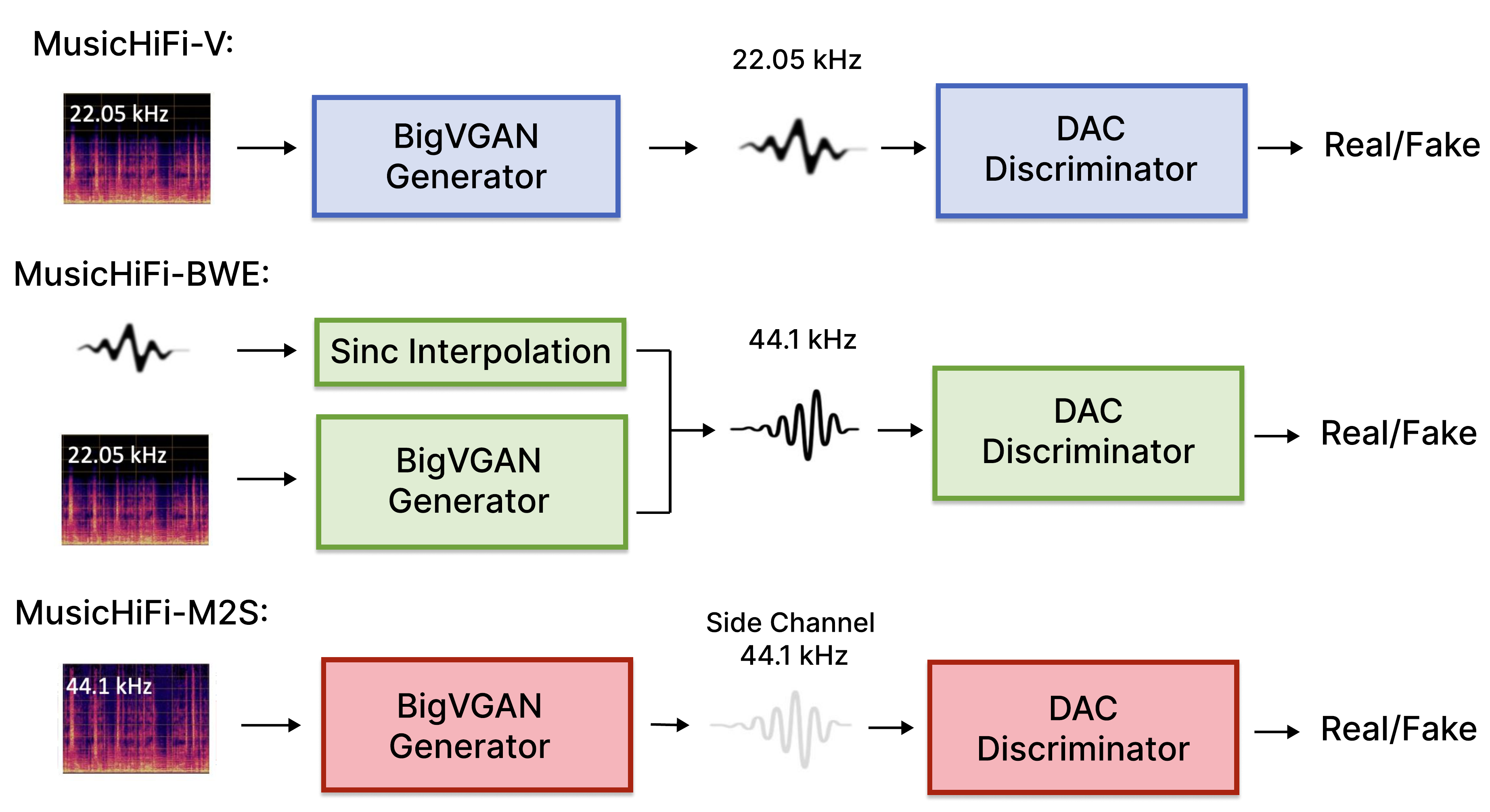}
    \vspace{-.05cm}
   \caption{}
   \label{subfig:diag2}
\end{subfigure}
\caption{\gz{(a) MusicHiFi inference via a cascade of our vocoder, bandwidth extension (BWE), and mono-to-stereo (M2S) modules that use a shared architecture, but with different weights. (b) MusicHiFi GAN-training for  our vocoder, BWE and M2S (top-to-bottom), separately. Our unified inference and training scheme enables novel, high-performing BWE and M2S.}}
\end{figure}

We propose MusicHiFi --- a new efficient high-fidelity stereophonic vocoder shown in~\fref{subfig:diag1}. 
Our method uses a GAN-triplet cascade that converts low-resolution single-channel mel-spectrograms (e.g., 22.05 kHz) to stereo high-resolution waveforms (e.g., stereo 44.1 kHz) via vocoder, BWE, and M2S modules. 
Our method can be integrated into mel-spectrograms music generators, used to enhance recording fidelity, and/or for spatialization. 
Compared to diffusion-based BWE methods~\cite{liu2023audiosr} and AR spatialization, our GAN-based method is much faster and differentiable. 
We evaluate our approach via objective and subjective listening tests and find our approach yields equal or better vocoding, BWE, and M2S quality with much  inference speed. Our contributions are:
\begin{itemize}
    \item A unified GAN-based generator, discriminator, and training recipe for vocoding, BWE, and M2S upmixing,
    \item A new fast BWE method that maintains low-frequency content in upsampled audio via skip connections, and
    \item A new fast mono-to-stereo method that uses a middle-side stereo encoding~\cite{johnston1992sum} that fully preserves monophonic content and offers superior spatialization width control.
\end{itemize}
\gz{By creating a unified approach to vocoding, BWE, and M2S, we achieve a significant improvement in audio quality, spatialization, and inference speed compared to past methods.}

\section{Background}
The most relevant recent works to our method include BigVGAN~\cite{lee2023bigvgan} and the Descript audio codec (DAC)~\cite{kumar2023high}. BigVGAN is a recently proposed vocoder method that has achieved state-of-the-art performance for generating high-fidelity waveforms from mel-spectrograms at speeds significantly faster than real time on a single GPU~\cite{lee2023bigvgan}.
The BigVGAN generator employs a neural network architecture composed of a stack of transposed 1D convolutions, each followed by an anti-aliasing multi-periodicity composition (AMP) block which internally uses a Snake activation function~\cite{ziyin2020neural}. Past work demonstrated AMP blocks are able to generate waveforms with fewer high-frequency artifacts and provide substantial improvements in both objective and subjective assessments. Furthermore, AMP blocks have been found to improve robustness for out-of-distribution vocoding and strong extrapolation ability.

\gz{DAC is a neural-based codec that follows the SoundStream generator~\cite{zeghidour2021soundstream} architecture with Snake activations, an improved GAN-based discriminator, an enhanced reconstruction training objective, and a residual vector quantization scheme that achieves state-of-the-art compression.}
When we look at the discriminator differences, both the BigVGAN and DAC use time-domain multi-period discriminators (MPD) to capture multiple periodic structures as well as spectral-domain discriminators, but DAC replaces the BigVGAN magnitude-only spectral discriminators with a multi-band multi-resolution complex spectrogram discriminators (MMSD) to enhance high-frequency prediction and mitigate aliasing~\cite{kumar2023high}. \gz{Compared to the reconstruction loss proposed by BigVGAN, DAC uses multiple mel-bins over multi-scale spectrograms to improve training stability and convergence speed.}

\section{Methodology}

\subsection{Overview}
We introduce MusicHiFi, a new vocoding method based on unified triplet-GAN cascade that progressively upsamples audio as shown in~\fref{subfig:diag1}. Our approach involves three stages, each being modular and useful for different applications. First, single-channel mel-spectrograms \gz{of low sampling rate} are transformed into waveforms of the same sampling rate using a vocoder (MusicHiFi-V). 
Then, waveforms \gz{of low sampling rate} are converted into \gz{waveforms with full-bandwidth} via our BWE module (MusicHiFi-BWE). 
Finally, the single-channel \gz{waveforms} are upmixed to stereo audio through our M2S module (MusicHiFi-M2S). 
At each stage, we use an identical generator architecture, discriminator architecture, training objective as shown in~\fref{subfig:diag2}. 
\gz{The main differences between the three modules are the varying input and output dimensions, and the inclusion of an additional skip connection from the input in the BWE module.}

In more detail, for all three of our generator stages, we adopt the BigVGAN transpose 1D convolution + AMP block generator architecture~\cite{lee2023bigvgan} that inputs a mel-spectrogram and outputs audio. For our discriminator architecture, we use the DAC discriminator~\cite{kumar2023high}. 
For our training objective, we also adopt the DAC reconstruction loss and adversarial loss and remove the codebook learning objective since our focus is high-fidelity audio synthesis. 
Our final training objective for our generator ($\mathcal{L}_G$) and discriminator ($\mathcal{L}_D$) are:
\begin{equation}
  \begin{aligned}
  \mathcal{L}_G &=\sum_{k=1}^{K}\bigg[\mathcal{L}_{adv}(G;D_k)+\lambda_{fm}\mathcal{L}_{fm}(G;D_k)\bigg]+\lambda_{rc}\mathcal{L}_{rc}(G),\\ 
  \mathcal{L}_D&= \sum_{k=1}^{K}\bigg[\mathcal{L}_{adv}(D_k;G)\bigg],
  \end{aligned}
  \label{eq:obj}  
\end{equation}
where we apply $K$ discriminators, $D_k$ denotes the $k$-th subdiscriminator from MPD or MMSD, $\mathcal{L}_{adv}, \mathcal{L}_{rc}, \mathcal{L}_{fm}$ represent least-square adversarial loss~\cite{mao2017least}, reconstruction loss and $L_1$-based feature matching loss, respectively and $\lambda_{rc}$ and $\lambda_{fm}$ represent corresponding loss weighting term.
Specifically, $\mathcal{L}_{rc}=\sum_{i}||\log{S_i}-\log{\hat{S_i}}||_1$, where $S_i$ indicates the $i$-th mel-spectrogram from a list of mel-spectrograms with different fixed resolutions.
$L_{fm}$ aims to minimize the distances between real and generated  features from discriminator layers~\cite{kumar2019melgan}. 

\subsection{MusicHiFi-V} 
\gz{Our vocoder inputs mel-spectrograms of low sampling-rate and outputs audio waveforms with the same sampling rate, and follows our unified generator, discriminator, and training recipe described above.}  
We note that the original BigVGAN training recipe exhibits instabilities and is susceptible to mode collapse when scaled up to larger models~\cite{lee2023bigvgan}. 
\gz{To enhance performance while stabilizing training, we double the input length of the audio samples originally defined in BigVGAN, reduce the number of convolution layers in the AMP blocks, increase the convolution channel width to 2048.}
Furthermore, this configuration approximately matches the floating point operations per second (FLOPS) of HiFi-GAN~\cite{kong2020hifi}.

\subsection{MusicHiFi-BWE} 
\gz{Our BWE module takes audio with a low sampling rate as input and outputs full-band audio and follows our unified generator and discriminator architectural design, and training objectives.}
For our generator architecture, however, we make two small, but significant changes. 
First, we compute an intermediate mel-spectrogram representation for the input audio with half the hop size used for the vocoder to double the sequence length and match full-bandwidth waveform output. 

Second, we add a residual or skip connection between the input audio signal of low sampling rate and the full-bandwidth audio output, with a sinc interpolation block in between that performs 2x upsampling. 
The residual connection enables our BWE generator to more easily generate low-bandwidth content and allows our BWE generator to focus model capacity on generating high-frequency content. The discriminator also operates on the higher sample rate, full bandwidth audio~\cite{eskimez2019speech,kumar2020nu}. Note, during preliminary testing, we experimented not using the residual connection which did not work and applying a low-pass (LP) filter to the generated waveform, which slowed training and did not enhance performance.

\subsection{MusicHiFi-M2S}
For our mono-to-stereo (M2S) upmixer, we follow our unified generator architecture, discriminator architecture, and training recipe for a third time. To create a stereo effect from a mono audio signal, however, we leverage a mid-side encoding~\cite{johnston1992sum} to convert stereo left and right signals into a summation channel (mid-channel) and a difference channel (side-channel)~\cite{johnston1992sum}. We then train our M2S module to input the mel-spectrogram of the mid-channel $M$ and output the side channel waveform $S$, where $M=\frac{L + R}{2}$ and $S=\frac{L-R}{2}$, $L$ and $R$ are the left and right stereo channel, respectively. Subsequently, we reconstruct left and right output channels via $L=M+S$ and $R=M-S$.  

As a result of using mid-side encodings, our method is downmix-compatible in that we can take a mono channel, upsample it to stereo, downsample back to mono, and recover the original mono channel perfectly. This is not the case with alternative methods which typically degrade results after repeated applications. Furthermore, we can also add a control mechanism to adjust the spatialization width by controlling the energy ratio between the side and mid-channels. We can do this by normalizing the mid-channel and side-channel energies to 0 decibels (dB) and then adjusting the mid/side energy ratio via $\hat{S} \xleftarrow{} \alpha S$, where $\alpha = 10^{\gamma/20}$ and $\gamma$ is a scalar factor in decibels. When $\gamma > 0$, there will be more side energy and when $\gamma < 0$, there will be less side energy.

\section{Experiment and Results}

\subsection{Training details}
We train all of our models using an internal dataset of 1800 hours of licensed instrumental music (stereo 44.1 kHz). For training, we randomly crop a sequence of 16,384 samples and then apply module-specific pre-processing. 
For our vocoder, we use channel averaging and downsampling to 22.05 kHz with STFT settings of a 1024-sample window, 256-sample hop size, and 128-band log-mel spectrogram. 
For our BWE module, we use the same pre-processing as the vocoder but halve the window and hop size. 
For our M2S module, we use channel averaging with identical STFT settings as the vocoder. 
We use the scalar weights $\lambda_{fm}=1$ and $\lambda_{rc}=360$ for the training objective in Eq.~\eqref{eq:obj} and train all modules with a batch size of 45 for 500k steps, and select the optimal checkpoint via the minimum validation multi-resolution STFT distance (STFT-D). 
Model sizes per stage are approximately 46M params and 55 GFLOPS for one second of audio. All other parameters follow from the open-source implementations of the BigVGAN~\cite{lee2023bigvgan} generator and DAC~\cite{kumar2023high} discriminator.

\subsection{Baselines}
For vocoding, we compare against BigVGAN~\cite{lee2023bigvgan} and HiFi-GAN~\cite{kong2020hifi} all trained on the same data and input features. Our retrained HiFi-GAN model has 14M params and take 52 GFLOPS for one second of audio. We also train a large  HiFi-GAN-large model with 1024 input channels with 55M params, while taking 211 GFLOPS for one second of audio. For BWE, we compare against Aero~\cite{mandel2023aero}, a recent state-of-the-art BWE method that uses a encoder-decoder architecture with LSTM and temporal-based attention layers with 19M params and takes 85 GFLOPS. We also compare a pre-trained checkpoint of AudioSR~\cite{liu2023audiosr} for BWE (no training code available). For M2S, we compare against a CPU-only open-source implementation\footnote{https://github.com/s3a-spatialaudio/s3a-decorrelation-toolbox}  of decorrelation based method~\cite{schroeder1958artificial,fitzgerald2011upmixing}, denoted as DSP, that divides the signal into transients, noise, and harmonics and decorrelates non-transient content.

\subsection{Objective evaluation}
\begin{center}
    \begin{table}[t]
\centering
\caption{Vocoder objective evaluation. Our vocoder module yields better results than baselines, but is mildy slower.}
\resizebox{\columnwidth}{!}{\begin{tabular}{cccccc}
\toprule
\textbf{Dataset} &\textbf{Method} &\textbf{Mel-D}$\downarrow$& \textbf{STFT-D}$\downarrow$ & \textbf{ViSQOL}$\uparrow$  & \textbf{RTF}$\uparrow$ \\
\midrule
\multirow{4}{*}{DSD100}&HiFi-GAN~\cite{kong2020hifi}& 1.09 & 0.65&4.47&\textbf{3488}\\
&HiFi-GAN-large~\cite{kong2020hifi}&1.06&0.60&4.48&3409\\
& BigVGAN~\cite{lee2023bigvgan}&0.94&0.41&4.61& 1818\\
&MusicHiFi-V & \textbf{0.87}&\textbf{0.33}&\textbf{4.67}&1786\\
\midrule
\multirow{4}{*}{FMA}&HiFi-GAN~\cite{kong2020hifi}& 1.09 & 0.64 &4.52&\textbf{3703}  \\
&HiFi-GAN-large~\cite{kong2020hifi}&1.04&0.57&4.56&3614\\
& BigVGAN~\cite{lee2023bigvgan}&0.94&0.41&4.62&1829\\
&MusicHiFi-V & \textbf{0.87}&\textbf{0.35}&\textbf{4.67}&1807\\
\bottomrule
\end{tabular}
}
\label{tab:vocoder}
\end{table}

\end{center}
\vspace{-0.8cm}
\begin{center}
    \begin{table}[t]
\centering
\caption{BWE objective evaluation for full-band audio. Low/high-band results are in parentheses.
* AudioSR outputs have a high-frequency EQ boost that causes evaluation issues.}
\resizebox{\columnwidth}{!}{
\begin{tabular}{cccccc}
\toprule
\textbf{Dataset} &\textbf{Method} &\textbf{Mel-D}$\downarrow$& \textbf{STFT-D}$\downarrow$ & \textbf{ViSQOL}$\uparrow$ & \textbf{RTF}$\uparrow$ \\
\midrule
\multirow{3}{*}{DSD100}&Aero~\cite{mandel2023aero}& \textbf{0.51} (0.05/1.16)	&0.12 (0.02/0.54) &\textbf{4.18}&19\\
&AudioSR*~\cite{liu2023audiosr}& 1.23 (0.64/2.25) & 0.51 (0.36/1.68) &3.54 &4\\
&MusicHiFi-BWE & 0.55 (0.08/1.18) & \textbf{0.11} (0.02/0.56) &4.14&\textbf{1639}\\
\midrule
\multirow{2}{*}{FMA}&Aero~\cite{mandel2023aero}& \textbf{0.89} (0.07/1.60) & \textbf{0.24} (0.03/0.73) &\textbf{4.12}&19\\
&AudioSR~\cite{liu2023audiosr}& 1.68 (0.68/3.18) & 0.68 (0.39/2.33) &3.25 &4\\
&MusicHiFi-BWE & 1.01 (0.09/1.76)& 0.26 (0.02/0.79) &4.08&\textbf{1613}\\
\bottomrule
\end{tabular}
}
\label{tab:bwe}
\end{table}

\end{center}
\vspace{-0.8cm}
\begin{center}
    \begin{table}[t]
\centering
\caption{Comparison of objective metrics for M2S system. Values in Mel-D and STFT-D include mid/side channels.}
\resizebox{\columnwidth}{!}{
\begin{tabular}{ccccccc}
\toprule
\textbf{Dataset} &\textbf{Method} &\textbf{Mel-D}$\downarrow$& \textbf{STFT-D}$\downarrow$ & \textbf{ViSQOL}$\uparrow$ &  \textbf{RTF}$\uparrow$ \\
\midrule
\multirow{2}{*}{DSD100-test}&DSP& 1.07/1.87	&1.09/1.70 &4.69&5 (CPU)\\
&MusicHiFi-M2S& \textbf{0.00/1.70}&\textbf{0.00/1.53} &\textbf{4.73}&\textbf{1539} (GPU)\\
\midrule
\multirow{2}{*}{FMA-small}&DSP&0.99/2.29 & 1.08/2.16 &4.70&4 (CPU)\\
&MusicHiFi-M2S& \textbf{0.00/2.03} & \textbf{0.00/1.88} &\textbf{4.73}&\textbf{1554} (GPU)\\
\bottomrule
\end{tabular}
}
\label{tab:m2s}
\end{table}
\end{center}
For objective evaluation, we use 673 clips from FMA-small~\cite{fma_dataset} and the test split of the accompaniment track from the DSD100 test dataset~\cite{SiSEC16}. 
For both test datasets, every segment has a duration of 30 seconds.
For objective evaluation metrics, we use a suite of four metrics including ViSQOL~\cite{hines2015visqol}, mel distance (Mel-D) and STFT-D.
ViSQOL is a perceptual quality metric that estimates a mean opinion score based on the spectral similarity to the ground truth.
The Mel-D and STFT-D measure the spectral distances between the reconstructed and ground-truth audio under a mel and a linear frequency scale, respectively.
We use the real time factor (RTF) metric or time processed over time elapsed on an A100 GPU to measure speed.

\subsection{Objective evaluation result}
Vocoder objective evaluation results are shown in Table~\ref{tab:vocoder}. 
We find our proposed method outperforms BigVGAN and HiFi-GAN on Mel-D, STFT-D and ViSQOL results while maintaining a lower RTF on both datasets. We also find the RTF of our vocoder method is lower compared to HiFi-GAN, but is still very fast and almost 2000x real-time on an A100. 

BWE objective evaluation results are shown in Table~\ref{tab:bwe}. We find both our proposed method and Aero yield comparable STFT-D, Mel-D, and ViSQOL results. When compared to AudioSR, we found that AudioSR is easily influenced by scale variations and also has a notable presence of high frequency components. 
In an effort to address this issue, we computed a scale adjustment factor by downsampling the generated waveform to 22.05 kHz normalizing the energy to the ground truth. Despite these adjustments, a significant gap remains in the objective metrics remains, likely from difference in training datasets~\cite{liu2023audiosr}. We also find the RTF of our BWE module is approx. 80-400x faster than alternatives.

M2S objective evaluation results are shown in Table~\ref{tab:m2s}. We find our method outperforms the DSP decorrelation method on STFT-D, Mel-D, and ViSQOL. Furthermore, it is important to note that the error of the mid channel is zero for our method since our M2S method is downmix compatible and maintains the original mid channel and only estimates the side channel. We also find the RTF of our BWE module is over 300x faster than the DSP method by way of efficient GPU compute.

\subsection{Subjective evaluation}
\begin{figure}[t!]
\centering
    \includegraphics[width=0.5\textwidth]{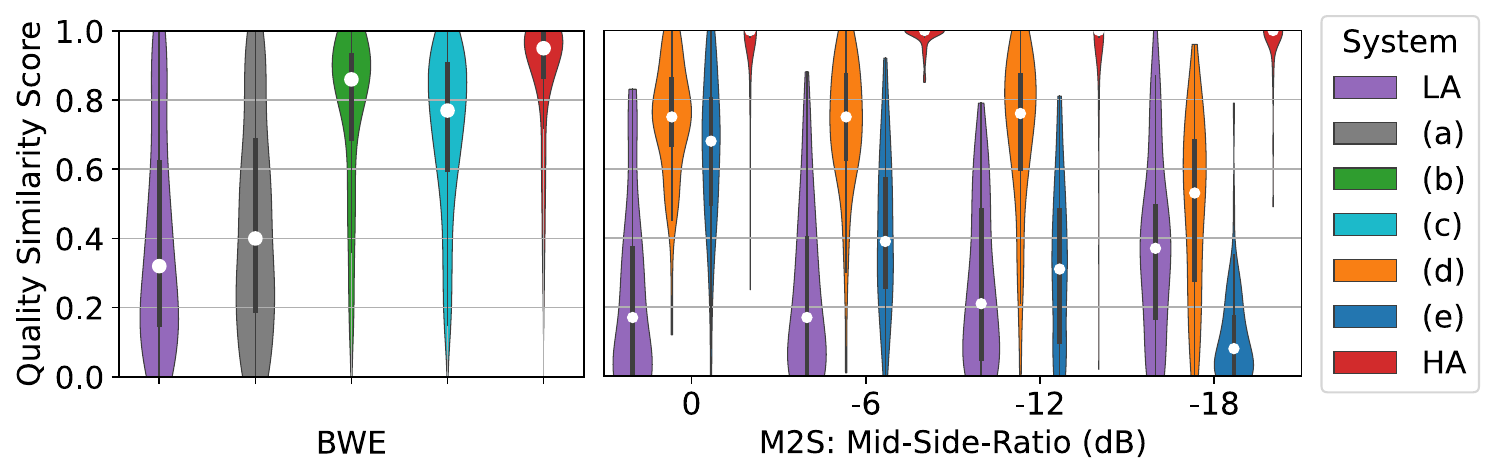}
    \caption{Subjective listening test violins plots. BWE test (left) and M2S test (right). Test conditions include (a) AudioSR, (b) cascaded MusicHiFi-V and Aero and (c) MusicHiFi-V+BWE (d) full MusicHiFi (e) MusicHiFi-V+BWE with DSP.}
    \label{fig:mushra_all}
\end{figure}

We performed two subjective listening tests to evaluate our BWE and M2S independently~\cite{itu2014recommendation}. 
For our BWE and M2S test, we recruited 20 and 23 participants with diverse audio backgrounds, respectively, and used a multiple stimuli with hidden reference and anchor (MUSHRA) protocol and the Web Audio Evaluation Tool~\cite{waet2015}. The goal of our BWE tasks was to have participants rate \textit{quality similarity} relative to ground-truth 44.1 kHz mono music. The goal of our M2S task was to rate \textit{quality similarity} relative to ground-truth stereo as well as test spatial controllability by varying the target mid-side energy ratio (i.e., from 0 to -18dB) given that performance varies heavily on the spatialization level. 

For our BWE listening test, we created six test examples from the FMA-small dataset, with each sample being 4 seconds long. Test conditions included (a) AudioSR, (b) cascaded MusicHiFi-V and Aero and (c) MusicHiFi-V+BWE as well as the LA and HA. The goal of this task is to understand the perceptual quality of comparable BWE algorithms. For the M2S evaluation, we prepared twelve listening samples, each 3 seconds long, selected from an internal test dataset instead of the FMA dataset as we found a large number of FMA clips were poorly spatialized. Test conditions included (d) the full MusicHiFi and (e) cascaded MusicHiFi-V+BWE with DSP for M2S comparison as well as the LA and HA.  All samples for the two tasks are loudness normalized to -23 dBFS before and after being fed into the cascaded methods. For both tests, 22.05 kHz mono signals were used as the low-anchor (LA) and a hidden reference as high-anchor (HA).

\subsection{Subjective evaluation results}

The results of our listening tests are shown in~\fref{fig:mushra_all}. 
When we compare the BWE subjective evaluation results, samples from (a) AudioSR ranks the least compared to other baselines. This result matches our earlier qualitative analysis that AudioSR has a strong high-frequency boost. 
We also find that (b) MusicHiFi-V + Aero ranks slightly above our BWE method, but we believe this is reasonable considering Aero uses an U-Net architecture with internal BiLSTM layers versus our convolutional architecture that is dramatically faster. 
We further conducted multiple post-hoc paired t-tests with Bonferroni correction~\cite{holm1979simple} for each condition vs. our method. We find there is no statistical significance between our method and Aero, while our method and Aero rank above AudioSR.

For M2S evaluation, we find that our (d) MusicHiFi performs best under different M/S panning coefficients and samples produced from the method (e) MusicHiFi-V+BWE with DSP perform similarly to ours when the energy ratio between mid and side channel is the same (0dB). The difference between our approach and the DSP baseline is statistically significant for side/mid rations 6, 12, and 18 via multiple post-hoc paired t-tests with Bonferroni correction~\cite{holm1979simple}.

For further evaluation, please find sound examples using mel-spectrograms extracted from real music and generated via a diffusion model~\cite{novack2024ditto} at~\url{https://MusicHiFi.github.io/web/}.

\section{Conclusion} 
We proposed a new efficient, high-fidelity stereophonic vocoding method named MusicHiFi. Our method works via a cascade of three GAN models that convert mel-spectrograms to low-quality audio waveforms, upsamples the low-resolution audio to high-resolution audio via bandwidth extension, and finally renders stereophonic high-resolution audio. Our method can be integrated into mel-spectrogram based music generators, used to enhance the fidelity of a low-resolution audio, and/or used to spatialize monophonic music. Compared to past work, we contribute a unified GAN-based discriminator and generator design, a new downsampling compatible BWE module, and a novel mono-preserving mono-to-stereo module. We evaluated our method using both objective evaluation and two subjective listening tests and found our method yields comparable or better vocoding and BWE results while outperforming comparable M2S methods, has better spatialization width control, and is extraordinarily efficient.

\section{Acknowledgement}
\zd{Z. Duan would like to thank support from National Science Foundation grant No. 1846184.}

\bibliographystyle{IEEEtran}
\bibliography{mybib}

\newpage

\end{document}